\documentclass[aps,pra]{revtex4}
\usepackage{amsm ath,latexsym,amssymb}
\newtheorem{theorem}{Theorem}

\newtheorem{lemma}{Lemma}
\newtheorem{definition}{Definition}

\def\duzomniejsze{<\kern-.7mm<}
\def\duzowieksze{>\kern-.7mm>}

\def\textbf#1{{\bf #1}}
\def\beq{\begin{equation}}
\def\eeq{\end{equation}}
\def\be{\begin{equation}}
\def\ee{\end{equation}}
\def\ben{\begin{eqnarray}}
\def\een{\end{eqnarray}}
\def\beqa{\begin{eqnarray}}
\def\eeqa{\end{eqnarray}}
\def\eea{\end{array}}
\def\bea{\begin{array}}
\def\s{\,\,\,\,\,\,}
\newcommand{\bei}{\begin{itemize}}
\newcommand{\eei}{\end{itemize}}
\newcommand{\bee}{\begin{enumerate}}
\newcommand{\eee}{\end{enumerate}}
\def\ra{\rangle}
\def\la{\langle}

\def\<{\langle}
\def\>{\rangle}
\def\blacksquare{\vrule height 4pt width 3pt depth2pt}

\def\dt#1{{{\kern -.0mm\rm d}}#1\,}
\def\ypodpis{\raise4mm\hbox{$\omega$}}
\def\qde{the PQC }

\def\qas{QAS}
\def\E{{\cal E}}
\def\M{{\cal M}}
\def\e{\epsilon}
\def\singlet{\psi_+}

\newcommand{\EE}{{{\mathbb E}}}
\newcommand{\proj}[1]{\ket{#1}\bra{#1}}
\newcommand{\bra}[1]{\langle #1 |}
\newcommand{\ket}[1]{| #1 \rangle}

\newcommand{\acc}{\textsc{acc}}
\newcommand{\rej}{\textsc{rej}}

\newcounter{protoline}
\newlength{\boxwidth}
\newlength{\bigboxwidth}

\newtheorem{proto_body}{Protocol}
\newtheorem{algo_body}{Algorithm}[section]

\newenvironment{protocol}[1]{ 
\setcounter{protoline}{0}
\begin{minipage}{\boxwidth}
\newcommand{\step}[1]
        {\addtocounter{protoline}{1} {\item [\ \ \ \ \arabic{protoline}:]{\small##1 } }}
\begin{proto_body}[ #1 ]
\ \\[0mm] 
\begin{description} }{\end{description}\end{proto_body}\end{minipage}}
\newenvironment{algorithm}[1]{ 
\setcounter{protoline}{0}
\begin{minipage}{\boxwidth}
\newcommand{\step}[1]
        {\addtocounter{protoline}{1} {\item [\ \ \ \ \arabic{protoline}:]{\small##1 } }}
\begin{algo_body}[ #1 ]
\ \\ 
\begin{description} }{\end{description}\end{algo_body}\end{minipage}}

\newcommand{\proto}[2]{\ \\ \fbox{ \begin{protocol}{#1}#2 \end{protocol}}\ \\}

\newenvironment{protocol_cont}[0]{ 
\begin{minipage}{\boxwidth}
\addtocounter{proto_body}{-1}
\newcommand{\step}[1]
        {\addtocounter{protoline}{1} 
          {\item [\ \ \ \ \arabic{protoline}:]{\small##1 } }}
\begin{proto_body}
\ \\ 
\begin{description} }{\end{description}\end{proto_body}\end{minipage}}

\newcommand{\QED}
\blacksquare

\newcommand{\pf}{{\it Proof:}}
\newcommand{\key}{\{y,z\}}

\begin{document}
\title{How to reuse a one-time pad and other notes on authentication, 
  encryption and protection of quantum information.
  }
\author{Jonathan Oppenheim$^{(1)(2)}$,  Micha\l{} Horodecki$^{(2)}$ }
\affiliation{$^{(1)}$
Racah Institute of Theoretical Physics, Hebrew University of Jerusalem, 
Givat Ram, Jerusalem 91904, Israel}

\affiliation{$^{(2)}$Institute of Theoretical Physics and Astrophysics,
University of Gda\'nsk, Poland}
\date{Monday, June 23rd, 2003}
\begin{abstract}

Quantum information is a valuable resource which can be encrypted
in order to protect it.  We consider the size of the one-time
pad that is needed to protect quantum information in a number
of cases.  The situation is dramatically different
from the classical case: we prove that one can
recycle the one-time pad without compromising security.
The protocol for recycling relies on detecting whether eavesdropping
has occurred, and further relies on the fact that
information contained in the encrypted quantum state cannot be 
fully accessed.
We prove the security of recycling rates when authentication of quantum states
is accepted, and when it is rejected.
We note that recycling schemes respect a general law of cryptography 
which we introduce relating the size of
private keys, sent qubits, and encrypted messages.
We discuss applications
for encryption of quantum information in light of the resources needed
for teleportation.  Potential uses include  the 
protection of resources such as entanglement and the memory of
quantum computers. We also introduce another application: encrypted
secret sharing and find that one can 
even reuse the
private key that is used to encrypt a classical message.
In a number of cases, one finds that the amount of
private key needed for authentication or protection is smaller than 
in the general case.

\end{abstract}
\maketitle

When encrypting classical information,
the only method that gives unconditional security is the
Vernam cipher or one-time pad.  Such a private key is a random string of 
correlated bits shared between two parties, who we shall
call Alice and Bob.  By
XORing \cite{xor} a message with the private key, Alice
can send a message to Bob which cannot be read by an
eavesdropper (Eve).  However, this is a rather expensive
protocol because classically, Alice and Bob cannot
securely increase the size of their private key without
meeting.
When they have finished using their private key, they may therefore be
tempted to reuse it.  
Of course, reusing the private key is highly insecure, 
and Eve will be able to exploit redundancies in messages
to learn the random string of the key and
gain information about the messages.  
It may be that no one is actually
eavesdropping on Alice and Bob, but because they have
no way of being certain, they should never reuse the key.

Just as classical information is valuable and may
need to be encrypted, quantum information is also
of value, and 
recently, it has been proposed that one may also want
to encrypt it \cite{boykin-qe,mosca-qe}.  
Protocols have also been
introduced to authenticate quantum information
\cite{dan-auth} [c.f. \cite{debbie-qvc}\nocite{curty-qa}], 
and the case of encrypting classical
data that has been encoded in quantum states has also been
considered \cite{dan-unclonable}.  

In \cite{boykin-qe} and \cite{mosca-qe} (c.f. also \cite{lo-ccc}),
it was shown that if the source to be encrypted is
a quantum message of $m$ qubits, then the size of the
private key for perfect encryption needs to be twice as large as in the 
classical case.  I.e. a correlated random string of
$2m$ bits is necessary and sufficient to encrypt the quantum message.
Such a protocol is called a Private Quantum Channel (PQC)  
and care should be taken to distinguish between 
the PQC where quantum information
is encrypted (using classical bits), and quantum key 
distribution \cite{bb84}, where
one uses the properties of quantum states to encrypt 
classical data.

An immediate question which arises in the context of 
the PQC, is what exactly does
it mean to encrypt quantum information?  In classical
encryption, one usually has the situation that Alice knows some
message which she wants to send to Bob, and Eve does
not know the message and wants to learn it.  In the encryption
of quantum information, the opposite can be the case: Alice
may not know the state being sent, and Eve may know it,
and want it in order to use it.
Additionally, if the state is unknown to Eve, then the
quantum information is
in {\it some} sense, already encrypted.  Eve
can get no more than one classical bit of information
from a single unknown qubit.  She cannot access the 
quantum information.  The more likely use of the PQC 
is therefore when Eve
knows something about the state, and Alice and Bob want
to make sure that Eve can not use it.  For example, the
state may be some valuable resource such as the memory of
a quantum computer or entanglement, and Alice and Bob want
to {\it protect} this resource from being used by an adversary.

Perhaps an easier to define scenario exists in the authentication
of quantum information.  Here, Alice wants to send Bob a
state in such a way that Bob can tell if the state has been
tampered with.  The protocols for a quantum authentication
scheme (QAS)  
are closely related to the PQC, and
therefore, much of our discussion will be related to them.
In fact, any QAS is necessarily a PQC \cite{dan-auth}.

Notwithstanding the necessity proofs of \cite{boykin-qe,mosca-qe}, 
in light of the fact that Eve can only extract one classical
bit of information from a qubit, one might wonder in what sense
the two classical bits are needed.  
In fact, we will show that one need not use up this classical
key - it can be recycled. 
When encrypting or authenticating quantum information, one can
break the cardinal rule of cryptography and reuse the
one-time pad.  

A strong indication that key-recycling of a classical key is
possible, was given in \cite{debbie-qvc}.  There, shared singlets, rather
than a classical key,  
were used to encrypt
quantum information.  It was then shown that by performing
distillation on the singlets, one could reuse them.  A
qualitative analysis indicated that this was due to the
fact that the encrypted information was quantum, rather than
the fact that the private key was quantum.  It was shown that
an attacker's Shannon information about a classical key could be bounded,
although it was unclear the extent to which secure 
recycling was possible, or whether correlations in different messages could
be learned even though the key might remain secret.

key-recycling makes use of one of the
fundamental properties of quantum information -- it cannot
be copied or cloned \cite{clone}.  What this means is that 
the encryption is such that Eve
cannot make copies of several messages in order to
compare them and deduce the values of the private key.
Eve could of course steal or tamper with
the quantum message, but this will result in a disturbance which can be 
detected using an error correcting code. 
This is in stark contrast to the classical case where 
Alice and Bob have no reliable way of knowing whether someone has successfully
eavesdropped.  
For these reasons, for an authenticated quantum channel using $2m+2s$ bits, 
(where $s$ is a security 
parameter), $2m+s$ of the bits can be securely recycled when the QAS
is accepted.  When the QAS is rejected, just under $m+s$ bits can be securely recycled.
For the PQC one has a similar rate of recycling.
In order to prove such recycling, one needs to show that (i) the probability
that Eve gains information about the private key can be made arbitrarily
small when the protocol succeeds, and (ii) that Eve cannot exploit
correlations in recycled keys in order to gain information about correlations
between states sent in successive uses of the channel. 
%
%

The fact that one can recycle the private key may seem surprising at
first, however, a hint that recycling is possible is given by
teleportation.
Instead of using a private
quantum channel, Alice could instead teleport \cite{teleportation} her state to Bob.
If the classical channel used in teleportation is authenticated, then
teleportation also provides authentication of the quantum state.  Both \qde  
and teleportation require similar communication resources -- both require
$m$ uses of a quantum channel to transmit quantum states (either
half-singlets in the case of teleportation or the quantum states themselves
in the case of \qde).
However, teleportation completely avoids using any private key at all.  Teleportation also
has the added advantage that there is less danger that the message will be lost
in transmission -- the quantum state passes directly to Bob, and there is
no way to tamper or destroy it.  One can therefore ask, what it is that
the private key buys us -- why is a private key is needed at all?
 In \cite{dan-auth} it was noted that
\qde  has the advantage that it is non-interactive.  i.e. two rounds
of classical communication are not needed.  The private key therefore
buys us a decrease in classical communication.  This seems like a rather
expensive trade-off, given that classical communication is usually 
considered to be a cheap resource, while a private key is usually
considered valuable (although physically, these resources are incomparable
in that neither can be converted into the other).

One might therefore ask whether there are many situations where it is
advantageous to use \qde over teleportation.  We will therefore,
in Section \ref{sec:sharing} give examples where \qde uses
as much or less resources than teleportation.  These include  
secure secret sharing, where we find that the related teleportation protocol 
also requires a private key.  \qde therefore consumes less
resources than teleportation (i.e. uses (i.e. no classical communication).
Furthermore, we prove the result that in the case of secure secret sharing using
teleportation, the classical message which needs to be encrypted
uses a private key which can also be recycled.  

We will then
discuss other cases where \qde is useful, namely for protecting quantum resources. 
For example, one can use quantum data encryption to protect entanglement.
We will also analyze protecting the memory 
of a quantum computer from
being stolen and used.
We find that often, the attacker will almost always 
have limited abilities which can be exploited to
use a one-time pad which is smaller than the $2m$ bit bound
of \cite{boykin-qe} and \cite{mosca-qe}.  This ability to beat the
$2m$ bound in specific cases is in addition to
our ability to reuse the pad\cite{distinction}.  
For protecting $n$ bits of entanglement,
we will see that an $n$ bit reusable pad is sufficient, while
for protecting the memory of a quantum computer, 
it might be possible to use a reusable pad of a size given by the error-correction 
threshold \cite{shor-fault,dorit-bo-thresh}.

Although some of the examples we give may be of a practical
nature, our primary motivation for studying \qde is because
it is interesting 
and raises many 
questions considering the nature of encryption and of quantum
information.  In particular, it allows us to gain additional
insight into teleportation by decoupling the
sending of qubits from their encryption.
We will also see that even with key-recycling, the PQC obeys
a general rule which we prove,  regarding the maximum increase
of a private key $\delta K$ as a function of sent qubits $\delta Q$ and
sent private messages $\delta M$.  Namely
\beq
\delta K \leq \delta Q - \delta M \s .
\eeq
We also discuss the thermodynamical nature of such a law.

We will discuss key-recycling in Section \ref{sec:recycle}, and prove
that it is secure.
Next, in Section \ref{sec:applications} we
discuss applications of the PQC, including encrypted secret sharing and the
protection of entanglement and other resources.  We conclude in
Section \ref{sec:conclusion} with a brief discussion.

%
\section{Private key recycling}
\label{sec:recycle}


Let us imagine that Alice wants to send to Bob, 
a state $\rho$ composed of $m$ qubits, and we consider the possibility
that the
potential adversary Eve may have some prior knowledge of the
states that will be sent, or may know from what distribution
they arise.  To create a PQC, Alice ``q-encrypts'' the state using her 
$2m$ bit private
key using the method of \cite{boykin-qe,mosca-qe}. Essentially, 
to each qubit, Alice conditionally
applies a bit flip in the z-direction, and then conditionally applies
one in the x-direction, using
two bits of her private key as the control bits.   The private key is a classical 
bit string randomly chosen from the uniform distribution ${\cal K}$.
The encrypted state $\rho_o$ is now 
maximally mixed regardless of what the initial state was, and
can be sent to Bob.  Such a procedure gives the private quantum
channel (PQC).

\begin{definition} 
 A secure Private Quantum Channel (PQC) with error $\epsilon$  
   is a set
  of classical keys ${\cal K}$  and computable super-operators
  $A_k$ and $B_k$ for each key $k$ such that:
  \begin{itemize}
  \item 
  For any
  ensemble $\E=\{p_i,\rho_i\}$, $A_k$ acts on all $m$-qubit states of $\E$
    and outputs an ensemble $\E'= \{p_i,\rho_{ik}\}$ 
     \item For all $\rho_i$ and keys $k\in{\cal K}$: $B_k
    (A_k(\rho_i)) = \rho_i$
  \item For all ensembles $\E$ and measurements $\M$ 
  acting on $\E'$, $H(v:i)\leq \e$ , where $H(v:i)$ is the mutual
information between measurement outcomes $v$ and the members
of the ensemble $i$.
\end{itemize}  
\end{definition}
In other words, a PQC is an encryption/decryption scheme, such that
for all ensembles, the probability that an eavesdropper learns more than $\e$ about
which state is being sent is small.  This definition is somewhat different
to that of \cite{mosca-qe, boykin-qe} in that we define the PQC with respect
to how much information can be gained by an eavesdropper
(as opposed to just requiring that PQC produce the maximally mixed state
$\rho_o$).
Note that the definition, though based on Shannon information,
does not refer to Eve's knowledge of the states.
Indeed, in the PQC paradigm, one assumes  that Eve may know the states
from the very beginning.
Note also that such a scheme (unlike teleportation) is only
one-way i.e. Alice just sends the encrypted state to Bob,
and no classical communication need be used.

We will discuss key-recycling in such a scenario.  However, before
proceeding to this discussion, we will first consider the case
of key-recycling in quantum authentication protocols, because there
 are less subtleties involved.

\subsection{Key-recycling in quantum authentication protocols}

It usually makes
little sense to encrypt quantum information without authenticating it in
some way, or protecting the state with some form of error correction.  
This is because unlike a classical message,
any eavesdropping on the channel may damage the state.  Since the states
may be unknown to Alice, the state can not be resent. 
A Quantum Authentication Scheme (QAS) is some encoding that Alice performs using a private key
shared with Bob, which enables Bob to tell whether the state
has arrived unaltered.  Bob will either accept or reject
the state,  depending on whether he believes he received
the correct state.
Again, like the PQC, the scheme is one-way.
\begin{definition} 
  \cite{dan-auth}
 A one-way \emph{quantum authentication scheme} (\qas) is a set
  of classical keys ${\cal K}$  and computable super-operators
  $A_k$ and $B_k$ for each key $k$ such that:
  \begin{itemize}
  \item $A_k$ takes an $m$-qubit state $\rho$ 
    and outputs a system $\rho_k$ of $m+s$ qubits.
  \item $B_k$ takes as input the (possibly altered) state
    $\rho_k'$  and outputs an $m$-qubit state $\rho'$, which includes 
    a single qubit which
    indicates acceptance or rejection (denoted 
    by $\ket{\acc},\ket{\rej}$).
  \end{itemize}
  
  \end{definition}

  A two-way QAS allows for communication between Alice and Bob
  during the protocol.  The scheme is secure if the probability
  that the protocol is accepted, and that the state is not $|\psi\ra$ 
  is less than $\e$.  I.e.

  \begin{definition} \cite{dan-auth}
\label{def:secure-qas}
    A quantum authentications scheme is secure 
 with error $\epsilon$ if for all states $|\psi\ra$ 
it satisfies:
  \item \emph{Completeness:} For all keys $k\in{\cal K}$: $B_k
    (A_k(\proj{\psi})) = \proj{\psi}\otimes \proj{\acc}$
  \item \emph{Soundness:} $Tr(P\rho')\geq 1-\e$ 
  where $P=|\psi\ra\la\psi|\otimes|\acc\ra\la \acc| +
  I_m \otimes |\rej\ra\la \rej|$ and $\rho'$ is again the output state.
    \end{definition}
 Here, security is defined for pure states, and can be extended to
 mixed states by the linearity of quantum mechanics.

We will consider a pessimistic scenario from the point of
view of key-recycling.  Namely, we imagine that Eve may know exactly
what state is being sent.
However, as in \cite{dan-auth} we will
assume that Eve has not already managed to acquire part of the state
\cite{debbie-qas-point}.
We will then see that Eve cannot learn too much about the private
key because she cannot access all the quantum information.

We will show that 
almost the entire key 
can be recycled when the protocol 
is accepted.  When the authentication protocol is rejected, we find
that
half the key can be recycled.   
We will show that Alice and Bob can place a sufficient bound on  
Eve's information (regardless of the initial state
and Eve's prior knowledge), 
that they can perform a process known as
privacy amplification \cite{privacy} to recycle their key.  
Essentially, they are
able to publicly communicate, to distill from their $2n$
bit key, a slightly smaller key  of which Eve has an exponentially
small probability of knowing anything about. We then show that this recycled key
can then be used in another round of the QAS. 

To get a true bound we will discuss a general authentication protocol 
based
on stabilizer purity testing codes (error correcting codes) used in Ref. \cite{dan-auth}.
 The protocol of this scheme (SQAS) is as follows
 \begin{figure}[h]
  \proto{Stabilizer based Quantum Authentication Scheme (SQAS)}
  {\label{pro:sqas}

    \step{Alice and Bob share a secret key $x$ of length $2m$ to be used
      for q-encryption.  For authentication, they additionally 
      agree on some stabilizer purity
      testing code $\{Q_z\}$ and two secret keys $z$ and $y$ of
      combined length $2s$.}

    \step{Alice uses $x$ to encrypt an $m$ qubit state $\rho$ as $\rho_o$.}
    
    \step{Alice encodes $\rho_o$ according to $Q_z$ for the code 
      $Q_z$ and adds syndrome $y$ to produce $\sigma$. This requires $s$
      additional qubits.  She then sends 
      the total state of $n=m+s$ qubits to Bob.}
  
    \step{Bob receives the $n$ qubits. Denote the received state by
      $\sigma'$.  Bob measures the syndrome $y^{\prime}$ of the code
      $Q_z$ on his qubits. Bob compares $y$ to $y^{\prime}$, and
      aborts if any error is detected.  Bob decodes his $n$-qubit
      word according to $Q_z$, obtaining $\rho_o'$. Bob q-decrypts
      $\rho_o'$ using $x$ and obtains $\rho'$. }  }
\end{figure}

Essentially, Alice not only q-encrypts the state $\rho$ using $2m$ bits
of classical key, but she also encodes the state using an error-correction
protocol based on a set of stabilizer codes,
and determined by some additional private key. An
additional $s$ qubits are used during transmission.   
The length $s$ is determined by the
degree of channel noise or extent of anticipated eavesdropping.  Such a
protocol can protect against an arbitrarily large amount of eavesdropping. 
An example of such a scheme was shown in \cite{dan-auth}
using a particular purity testing 
   code $\{Q_z\}$ which gave a soundness error $2n/s(2^s+1)$ using an 
   additional key of length $s +\log_2(2^s+1)$.
If the authentication scheme passes, we will see that they can 
recycle the key because
the probability that Eve has any information about the state or the key
can be made arbitrarily small, and Eve is product with the sent states.

  The proof of security of Protocol \ref{pro:sqas} given in
  \cite{dan-auth} is analogous to the
      Shor-Preskill security proof of BB84 \cite{shor-preskill}.  
      One essentially
      shows that such a scheme can be converted into a teleportation
      protocol without loss of security.


%

We now turn to proving security of key-recycling.  By security, we
mean that if one has two secure QAS with two keys $k$ and
$l$,  to authenticate two states, then the scheme is still secure
if one can replace the key $l$ with a key recycled from $k$.
Denoting the length of a key by $||$, we have  

\begin{definition} A Secure Key Recycling scheme of error $\epsilon$
  and efficiency $k'/k$ 
  is a map $\Lambda:{\cal K} \rightarrow {\cal K'}$ such that if $\E_k$ and
  $\E_{l}'$ are 
  secure QAS on Hilbert spaces  ${\cal H}$ and ${\cal H'}$ then when
the QAS are accepted
 $\E_{k'}'\E_k$ is a secure QAS on  
   ${\cal H}\otimes{\cal H'}$ with error $\e$  when $|l|=|k'|$ with 
 $k'=\Lambda(k)$.
\end{definition} 
The error $\e$ depends not only on 
the key-recycling scheme
but also on the security of the QAS.

The proof in \cite{dan-auth} of security of the QAS assumed that Alice
and Bob are authenticating pure states (or parts of pure states).
I.e. it is assumed that the adversary does not already hold part of
the state being sent.  After the arXiv version of this paper appeared, this condition
was removed in the more general and detailed work of \cite{HLM-recycle}.
We will also make this assumption,
and under it, now prove the following

\begin{theorem} \label{thm:main} For the QAS of Protocol \ref{pro:sqas} 
there exists a secure key recycling scheme with efficiency rate
$\frac{2m+s}{2m+2s}$.
   \end{theorem}

To prove Theorem \ref{thm:main} 
we first need the following lemma saying that 
sending half of a singlet does not need encryption by $x$.

\begin{lemma}\label{lem:me} Consider the maximally entangled state 
  composed of two subsystems $\ket{\psi_+}=\sum_i|i\ra_1|i\ra_2$ on
  ${\cal H}_m\otimes{\cal H}_m$, with dimension $2^m\times 2^m$ and $\rho$ being
  the state obtained from tracing out one of the subsystems.   
  Then a secure SQAS using Protocol \ref{pro:sqas} for $\rho$
  with a key of length $2m+2s$ is also secure using a key of length $2s$.
  \end{lemma}

  In other words, for half of a maximally entangled state, the key $x$ is
  not needed, only the key $y$ and $z$.
  
  \pf\ Alice uses Protocol \ref{pro:sqas} on subsystem 2 except that
  neither step 2, nor the key $x$ is used.  We then note that because
  \beq
  I\otimes U \ket{\psi_+}=U^*\otimes I\ket{\psi_+}
  \label{eq:starinvar}
  \eeq
  Alice can replace step 2 by
  q-encrypting subsystem 1.  Then, due to Eq. \ref{eq:starinvar} this
  protocol is completely equivalent to Protocol \ref{pro:sqas} and is a
  secure QAS.  Then, since Alice's actions on subsystem 1 commute with Eve's and Bob's actions on 
  subsystem 2, it cannot matter whether Alice actually performs the
  q-encryption of subsystem 1. \QED

Now, we prove Theorem \ref{thm:main}, using a technique analogous to that
      of Shor-Preskill.  
      
\pf\
Let us consider a variation of Protocol \ref{pro:sqas} acting on a pure
state $\psi$ 
by assumption
 (by the linearity of quantum mechanics, our proof will hold if
 only part of the state is being sent.  I.e. it holds for any
 $\rho$ as long as the purification is not with Eve).  
The {\it modified QAS}, Protocol \ref{pro:mqas}, differs in step 2, and the
      final step from Protocol \ref{pro:sqas}.

 \begin{figure}[h]
  \proto{modified QAS}{\label{pro:mqas}

    \step{Alice prepares the state 
${1\over \sqrt 2^{2m}}\sum_x |x\ra$ where the $|x\ra$
      are $2^{2m}$ orthogonal states.      
      For authentication, Alice and Bob additionally 
      agree on some stabilizer purity
      testing code $\{Q_z\}$ and two secret keys $z$ and $y$ of
      combined length $2s$.}

    \step{Alice q-encrypts $\psi$ conditionally on $|x\ra$ 
      creating the state ${1\over \sqrt 2^{2m}} \sum_x|x\ra|\psi_x\ra$}
    
    \step{Let us call the subsystem which is the mixture of $\psi_x$'s
      $\rho$.
      Alice encodes $\rho$ according to $Q_z$ for the code 
      $Q_z$ with syndrome $y$ to produce $\sigma$. This requires $s$
      additional qubits.  She then sends 
      the total state of $n=m+s$ qubits to Bob.}
  
    \step{Bob receives the $n$ qubits. Denote the received state by
      $\sigma'$.  Bob measures the syndrome $y^{\prime}$ of the code
      $Q_z$ on his qubits. Bob compares $y$ to $y^{\prime}$, and
      aborts if any error is detected.  Bob decodes his $n$-qubit
      word according to $Q_z$, obtaining $\rho'$. 
     }  
     
     \step{If Bob accepts, Alice measures $|x\ra$ to obtain a random string
     $x$}
     }
\end{figure}

      First we note that because $\psi$ has essentially been q-encrypted
      by the action's of Alice, $\rho$ is the maximally mixed state $\rho_o$.
      Therefore, the state $\sum_x|x\ra|\psi_x\ra$ is a maximally entangled
      state $\psi_+$ of $2m$ qubits (Actually up to local unitaries, on Alice's side 
      it is half a singlet tensor the initial state).
Therefore, by Lemma \ref{lem:me}, if Bob accepts,
      then Alice and Bob share a  state $\rho'_{AB}$ that has an overlap with 
$\psi_+$ equal to $1-\e'$. One can now use standard reasoning originating in 
\cite{ekert-qkd}: since Alice and Bob share an (almost) pure state, they are decoupled 
from Eve. 

      Alice can then measure $x$ without Eve learning
      what $x$ is.  From the point of view of Eve, Protocol
      \ref{pro:mqas}  is completely equivalent to Protocol \ref{pro:sqas},
      and therefore, if Bob accepts the authentication, the probability that
      Eve has 
obtained more than $\e$
      information about
      $x$ is small.

      It should be pointed out that Protocol \ref{pro:mqas} is not
   equivalent to \ref{pro:sqas} from the point of view of Bob, since he does
   not have the key $x$.  However, this is not relevant, since he decides
   whether to accept the authentication before decoding using $x$.  Therefore
   neither his actions nor Eve's depend on the modifications that Alice
   makes.  One can in fact show that the preceding protocol is in fact, 
   equivalent to teleportation.


Let us now prove that Eve is product with $x$ and the sent state. 
This will ensure not only reusability of the key in another QAS, 
but in any other cryptographic task according to \cite{Ben-Or-Mayers} and 
\cite{BHLMO}.
This  essentially
follows from the fact that if the authentication is accepted, the
protocol is equivalent to teleportation.
Let us consider the total state of the system, before step 5
of Protocol \ref{pro:mqas}.  Using the fact that all ancillas and all
actions by the parties can be represented as unitaries acting on pure
states, we can write the total state of the system in the form
\beqa
\Psi &=& \frac{1}{2^m2^s}\sum_{x,y,z}
\ket{x}U_BU_EU_A(\ket{\psi_x}\ket0_E\ket{0}_{A'} \ket{y,z}_{A''}\ket{y,z}_{B''})\ket{y,z}_R
\nonumber\\
&=&
\sum_i\sqrt{p_{i}}\ket{\psi_i}\ket{\phi_{i}}
\eeqa
Here, the key used in the authentication is encoded in the states  
$\ket{y,z}_{A''}\ket{y,z}_{B''}$ which is labeled 
by the particular $2s$ bit key $\key$ that is used in the QAS.
Since this is a classical key, we purify it on an imaginary reference system  
$\ket{yz}_R$ to which none of the parties have access.
The first unitary $U_{A}$ is the further encoding done by Alice, and it acts
on the states $\psi_x$ and the $s$ qubit ancilla $\ket{0}_{A'}$
conditional on the authentication key $\ket{y,z}_{A''}$.
The second $U_E$ are the actions of the eavesdropper acting on the $m+s$ qubits sent across 
the channel, and $U_B$,
the decoding of Bob, acting on the ancilla and sent 
state $\psi_x$ (conditioned on $\ket{y,z}_{B''}$).  
 
In the second line we have rewritten $\Psi$ in terms
of the eigenbasis of maximally entangled states (one of which is the state 
$\singlet\equiv\psi_0=\sum_x \ket{x}\ket{\psi_x}$), correlated with states on the ancillas $\phi_i$ (which need
not be orthogonal).
Now, if the QAS is accepted, then we have according to Definition 
\ref{def:secure-qas} of a secure QAS
\beq
Tr(Tr_{A'A''B''RE}(\proj{\Psi})\proj\singlet)\geq 1-\e
\eeq
which implies that with probability $p_0 \geq
1-\e$, the state of the system is $\ket\singlet\otimes\ket{\phi_0}$
which implies that with arbitrary high probability, Eve is product with the key 
$x$ and the encoded state $\psi_x$ (even more strongly, all ancillas, 
as well as the key $\key$
are product with $x$ and the sent state $\psi$). 
From the composability theorem \cite{Ben-Or-Mayers}, this implies that the key $x$ can
be reused in the second QAS (using the same methodology of \cite{BHLMO}, since one can also regard
this protocol in terms of security of key).  

Note that we not only had to proves Eve's lack of knowledge about $x$, but  
also that she was decoupled from the sent states.  This then gives
that the recycled key $k'$ can be used in a secure QAS.  The danger,
is that $k'$  might not be reusable, because it is correlated to the
sent states.  For example, in Sec. \ref{ss:repqc}
we will see that with respect to the PQC, one must be concerned
with the fact that
Eve could learn about correlations between
different sent states, even though she learns nothing about the key.
However, reusability of $k'$ in the QAS 
follows from the definition of the QAS.  Essentially, since the QAS
is not defined in terms of Eve's knowledge, but in terms of Bob
verifying
he got the correct state, one needn't worry whether Eve is
correlated
with the sent states - we can assume she already knows them.

Let us now turn to the key $\key$ which was used in the purity testing code.  
It is of length $2s$, and we now show that we can perform privacy amplification
to obtain a key of length $s$ which can also be re-used.  In the first arXiv version
of this paper, we showed that the probability that 
Eve obtains more than an exponentially small amount of information about the key is exponentially small,
using results of \cite{nayak98,nayak99,konig}. And further, that this part of the key was product with
the sent states. 
Since then, the results of \cite{RennerK04-key} appeared, and we can use their stronger
bounds to show that Eve is in fact, product from $s$ of the bits which can be recycled from 
the $2s$ key used in the authentication step. Thus we can formally prove
that this part of the key can also be re-used using the stronger security 
definition of composability.
%
%


Alice and Bob will need to perform privacy amplification \cite{privacy}
on the key $\key$ to produce a smaller key of length $s$, and in order
to prove the security of this privacy amplification, we use
%
%
\cite{RennerK04-key}:

  \begin{lemma}\label{le:me}
     Given a set of two-universal Hash functions \cite{hash} $G$ from
$J$,
a random classical distribution with density matrix $\rho_J$, to 
a distribution $T$ with range $\{0,1\}^{t}$ 
and density matrix $\rho_T(G)$, joint density matrices $\rho_{JE}$,$\rho_{TE}(G)$, and
marginal distribution $\rho_E=tr_J\rho_{JE}=tr_T\rho_{TE}(G)$
then
$\EE_G (tr|(\rho_{TE}(G) -\rho_T(G)\otimes\rho_E|)
\leq 2^{\frac{1}{2}(\log\lambda_{max}(\rho_{JE})+\log(rank(\rho_E))+t)}+2\epsilon$ 
where $\EE_G$ is the expectation value over $G$, and $\lambda_{max}$ is the largest eigenvalue.
    \end{lemma}
Intuitively speaking, the above lemma places a bound on how close the privacy amplified distribution is to being
product with an eavesdropper.
 
In the case under consideration, if the protocol is accepted, then as shown above, with probability $p\geq 1-\epsilon$ 
the total state of the system
is  $\ket\singlet\otimes\ket{\phi_0}$ with $\ket\singlet$ with Alice and Bob, and $\ket{\phi_0}$
on $A'A''B''R$ and $E$.  With high probability, Eve is completely product the $m$ qubit singlet, and the total state is
\beq
\ket{\singlet}_{AB}\otimes\ket{\phi_0}_{A'A''B''ER}\s .
\eeq
Upon acceptance of the protocol, we can thus simulate the situation  by giving
the entire $A'$ to Eve, since this can only further degrade the security.  Then,
with probability  $p\geq 1-\epsilon$, 
Eve's state $\rho_E$ has maximal rank $s$, since she is effectively only acting on the $s$ qubit ancilla.  
The largest eigenvalue $\lambda_{max}$
of the density matrix on $A'A''B''E$ is 
no greater than $1/2^{2s}$ (since the probability of any ${y,z}$ is this large.  Thus by
Lemma \ref{le:me}, $E_G (tr|(\rho_{TE} -\rho_T\otimes\rho_E|)$ will be small if we choose
the recycled key of length $t=s-2$.  Here, $\rho_T(G)$ is the density matrix of the privacy amplified
key $\key$.  This is precisely the condition needed for composability of a key found in \cite{BHLMO},
and ensures that just under half of the $2s$ bit key used in the authentication process can be recycled. 
%
The total efficiency of the recycling is then a rate of $(2m+s)/(2m+2s)$. 
\QED

\



Here some remarks concerning our use of composability are in order.
Our primary goal is to prove that the key can be recycled. 
In the case of $x$ we had two problems we fought with:
not only 1) Eve should have  small knowledge about 
the recycled key, but also 2) Eve should not be correlated with 
the total system: message plus recycled key.
This is precisely a place, where recycling of the classical one-time pad fails:
knowing that the key was used once more, Eve will get to know information about 
correlations between subsequent messages. 

In our case we assume from the very beginning that 
Eve knows the message. Then,  correlations of Eve with system "recycled key plus message"
would be dangerous because 
Eve would then get to know the key, 
and in some further round, she would break authentication 
(i.e. tamper with the sent state, without causing Bob's rejection).
Thus, both in case of key $x$ and $z,y$ we showed 
that the recycled keys are not correlated with the states 
subjected to authentication.

Let us now consider the case that the QAS is rejected.  
Typically, Bob will reject the authentication if the noise level
or eavesdropping is higher than anticipated.  I.e., if Alice uses an error
correction code which is not large enough. 
In such a case, we will show
\begin{theorem}\label{thm:rej}
For the QAS of Protocol \ref{pro:sqas}, if the protocol is rejected, 
there exists a recycling scheme with efficiency rate $\frac{m+s}{2m+2s}$
\end{theorem}

In the initial arXiv version of this paper, our recycling scheme in the case of rejecting
the QAS yielded $s$ bits of key.  We obtain a better rate of recycling by
simply giving Eve all the $m+s$ qubits.  Then, the maximal rank of Eve's state is thus
$m+s$, and the maximum eigenvalue of the total state (authentication key plus Eve's state) 
is bounded by $1/2^{2m+2s}$.  Thus by use of Lemma \ref{le:me} we can re-use the key
if we hash the total key to a size just under $m+s$.
\QED


An important aspect of the key-recycling scheme is that it doesn't
particularly take away from the advantage of the QAS over teleportation.
The QAS has the advantage that it is a non-interactive (only one-way
communication is needed, but at the price
of a private key).  Key-recycling will require a
small amount of back-communication (Bob must tell Eve whether he 
accepts or rejects the state).  However, the back-communication
and key-recycling can be done at the convenience of the two parties.
I.e. if Alice needs to send states so that Bob can use them right away,
then this can be done, and later, during a break in the transmission,
they can engage in key-recycling.  Therefore, a QAS plus key-recycling
still has an advantage over teleportation in terms of interaction.

\subsection{The slippery slope towards teleportation}

Having determined that the key can be recycled in QAS, one must then
wonder if it is needed at all.  Indeed, this is the case.
At the end of Protocol \ref{pro:mqas}, Alice after measuring what
$x$ is, can just tell Bob
the result publicly.  This is no less secure than 
Protocol \ref{pro:mqas} since the stabilizer code ensures that
Eve didn't touch the state, and therefore, by  the time she
learns what $x$ is, it is too late.
In fact, the keys $z$ and $y$ in Protocol \ref{pro:mqas} can
also be replaced by public communication from 
Alice to Bob (as long as the classical public communication
channel is authenticated).  This results in Protocol \ref{pro:tele}.
  
 \begin{figure}[ht]
  \proto{modified QAS}{\label{pro:tele}

    \step{Alice chooses random strings $x$, $y$ and $z$. 
      Alice and Bob additionally 
      agree on some set of stabilizer purity
      testing codes $\{Q_z\}$
      }

    \step{Alice q-encrypts $\psi$ conditionally on $x$ 
      creating the state $\psi_x$}
    
    \step{Let us call the subsystem which is the mixture of $\psi_x$'s
      $\rho$.
      Alice encodes $\rho$ according to $Q_z$ for the code 
      $Q_z$ with syndrome $y$ to produce $\sigma$. This requires $s$
      additional qubits.  She then sends 
      the total state of $n=m+s$ qubits to Bob.}
  
    \step{Bob receives the $n$ qubits. Denote the received state by
      $\sigma'$.  He indicates receipt to Alice.  }
    
    \step{Alice tells Bob $z$, and Bob measures the syndrome of $Q_z$ on
    $\sigma'$ obtaining the result $y^{\prime}$.
      Alice and Bob compare $y$ to $y^{\prime}$, and
      abort if any error is detected.  Bob decodes his $n$-qubit
      word according to $Q_z$, obtaining $\rho'$.  }  
     
     \step{If the protocol is accepted, Alice tells Bob  $x$ and Bob
    decrypts $\rho'$ to obtain $\psi'$}
     }
\end{figure}

That this protocol is secure can be seen by noting that it is 
essentially teleportation (as was Protocol \ref{pro:mqas}).  
In teleportation, Alice sends
$n$ half-singlets to Bob, and then they measure the syndrome $y$
of the random code $Q_z$.  If they both get the same $y$, they
can presume that the singlets are pure, and they begin teleportation.
The preceding protocol differs from teleportation, only in
that it is as if Alice has made the Bell measurement to start 
teleportation, and measured the syndrome of $Q_z$, before sending the
state to Bob.  Since such measurements don't change the density
matrix, they clearly can't reduce security.

This procedure, has the advantage over the QAS in that less
key is needed -- not a surprise given that much of it can be reused in
the QAS.  Of course, it is interactive, while for the 
QAS with recycling, the interaction can be performed at any time,
well after Bob has used his received state.  This in some
sense highlights the role of the secret key $x$ used in the QAS.  It's
purpose is not as much for encryption (since Alice could
use a random string instead and later tell Bob), but rather,
it serves a role of communication.  In other words, it allows
one to have a non-interactive protocol.  In fact, there
is still interaction -- the key was
distributed at some point in the past, and might be recycled
at some point in the future -- but this interaction can occur
at more convenient times.  The price (the secret key), is of
course rather high.



\subsection{Key-recycling for the Private Quantum Channel}

\label{ss:repqc}

Having discussed key-recycling for the QAS, we now turn to
key-recycling for the PQC.  We consider here a PQC such as
that of \cite{mosca-qe,boykin-qe} -- namely, for each qubit
of message, one uses 2 bits of private key, with Alice performing
one of four operations on each qubit conditional on the key.
We will now explain why to get key-recycling for the PQC,
one must actually use authentication or purity testing.  One
therefore must actually do key-recycling of the QAS.

At first, it might seem that for the PQC one can easily recycle  
half the key,
based simply on the generalized Holevo bound given in the Appendix.  
Eve will be ignorant of half the key, no matter
what states are being sent and how much she knows about the
states that are being sent.  

There is however, a problem with this.
Eve may not be interested in learning the key or the identity
of a particular state, but may
instead be interested in learning about correlations between the various
quantum states being sent between Alice and Bob.  
Let us imagine that Eve does not try to
learn anything about the state $\rho$ encrypted with $k$
but instead, just steals it.  
Then, when Alice and Bob send a new state $\rho'$ encrypted
with $k'$ she steals that as well.  Now she may not know much
about these two states, but she will learn something by
having both of them.  The reason
is that $k'$, although unknown to Eve, is in fact correlated
with $k$.  This means that Eve will be able to learn something
about the correlations between $\rho$ and $\rho'$, even though
she has no knowledge about $k$ and $k'$.

The amount of information that Eve can learn about
the key, depends not only on how much she knows {\it a priori}
about the states being sent, but also on which states being sent.  
For example, for
mixed states, from
the Holevo bound one has
that Eve will learn less than $m$ bits of information about
the key.
Even for pure states, Eve will in general learn much less.
Consider for example the state
\beq
|\pi/4\ra=\cos(\pi/8)|0\ra+\sin(\pi/8)|1\ra \s .
\eeq
It is easy to verify that this state gets encrypted 
into one of four non-orthogonal states $|0\ra$, $|1\ra$,
$|0+1\ra$, or $|0-1\ra$, conditional on the private key.  
These are the same states used in BB84\cite{bb84}.

Now, it may be that Eve did not steal the state, or
was unsuccessful in her eavesdropping.  It therefore seems a pity
for Alice and Bob to throw out $k$ if this is indeed the case.
Classically, Alice and Bob have no way to determine whether
eavesdropping was successful, but this is not the case for the
encryption of quantum information. Alice and Bob can indeed 
discover whether Eve tampered with their state using 
a QAS 
or
a purity-testing protocol 
to ensure that  they still possess the original state $\rho$.  Then,
because of the no-cloning theorem\cite{clone}, they
can be sure that Eve did not steal or tamper with $\rho$.

While an authentication scheme such as that of Protocol
\ref{pro:sqas} will be enough to enable key-recycling,
one can also have Alice publicly announce the authentication
strings $y$ and $z$, exactly as was done for Protocol \ref{pro:tele}.  
However, since these have to be done over 
an authenticated classical channel, it is doubtful that this would
be less resource-intensive.   

 
%


In summary, if one wishes to perform key-recycling of a PQC,
one needs to augment the PQC with some authentication.  The amount
of key-recycling is then given by Theorem \ref{thm:main}.  Since
any authentication scheme is also a PQC \cite{dan-auth}, the
theorem also guarantees security of the PQC.
As was noted in our proof of Theorem \ref{thm:main}, Eve will
be virtually product with the encrypted states $\psi_x$.  
It is for this reason that Eve is not able to gain information
about correlations between the various states being sent.

%
\subsection{The basic law of privacy}

It might make one uncomfortable that the length $K$ of the private key need not decrease
by $1$ for each qubit of encrypted message sent.  We now that classically, one cannot increase
privacy through communication.  However, we know that due to quantum key distribution,
the size of a private key can increase for each qubit sent.  Quantum communication therefore
allows one to increase the size of a private key by $1$ per qubit sent.  If the number of sent
qubits is $\delta Q$, and the size of an encrypted message which gets sent (whether quantum or classical) is
$\delta M$, then we note
the following general law.  For any communication between two parties
\beq
\delta K\leq \delta Q-\delta M
\label{eq:basic}
\eeq
where $\delta K$ is the change in length of the private key.
Such a law, while probably known on some intuitive level, has never been stated or proven to the
best of our knowledge.  We note that our recycling protocol respects such a law, as does teleportation.

\noindent {\it Proof:} Assume for contradiction that the basic law is violated.  Then, in the case where the 
additional message $\delta M$ is classical, we imagine that the message sent is used to create 
additional key of length $\delta M$ (since the private communication can always be communication of the private key).  Thus
a violation of the basic law would imply $\delta K \geq \delta Q$.  That the latter inequality can't be true
follows from \cite{pptkey}, where the theory of privacy was recast in terms of entanglement theory, and thus
it was shown that the relative entropy of entanglement, 
\beq
E_r(\rho_{AB})\equiv\min_{\sigma_{AB}\in sep}Tr(\rho_{AB}\log\rho_{AB}-\rho_{AB}\log\sigma_{AB})
\eeq
%
(where the minimum is taken over separable states) is an upper bound on the key rate.  
Since $E_r$ is less than $E_c$ (the entanglement cost) of a state (which is by
definition $Q$), it follows that  $\delta K \leq \delta Q$.  This gives the desired contradiction.

If the additional message $\delta M$ is quantum, then the quantum channel must work for all states (since by definition
of a quantum message the sent states are unknown and decoupled from the environment).  We could thus send half of a singlet
for each qubit of message, and then use it to create one bit of key.  We would then again have $\delta K \geq \delta Q$.
\QED

It is actually quite surprising that \cite{pptkey} is needed for what appears to be such a basic and simple result.
One imagines that  $\delta K \leq \delta Q$ would follow from the Holevo bound, but because a key is a strictly 
weaker resource than communication, and can be distilled using many rounds of public communication, standard techniques
appear not to work.

The laws of privacy bear a resemblance to thermodynamics - not surprising
given their information-theoretic nature.
 Alice and Bob
each possess a random string of maximal entropy.  $K$ therefore, represents a decrease in the total
entropy of Alice and Bob's string, due to correlations -- it is a {\it negentropy}.
In the absence of eavesdropping, both teleportation, and our recycling protocol
have $\delta K=0$, and are therefore optimal.  They are, in a sense,
      isentropic processes.
On the other hand, eavesdropping disturbs the system, and results in an increase of entropy
between Alice and Bob.  The size of the private key (or negentropy), goes
down.  This is like the second law of thermodynamics.
The leakage of information to Eve is  physically very closely related to the
processes that produce an increase of entropy in thermodynamics: 
Indeed, the process that conserves energy and
increases entropy is exactly eavesdropping I.e. it is pure decoherence (if
we will not consider just coarse graining).
In the case of no eavesdropping, then the bound of Equation
(\ref{eq:basic}) gets saturated, and you have a law which looks very
similar to the first
law of thermodynamics.  In that sense, one should think of the sent
messages as work which is being performed.

\section{Applications}
\label{sec:applications}


As noted in the Introduction, if two parties are interested
in sending an unknown state between them, they are likely
to chose teleportation over the PQC.  To teleport an unknown state,
Alice sends a number of halves of singlets to Bob, and then
they use distillation to ensure that they are indeed sharing pure
singlets (since the sent entanglement may get corrupted
due to noise, or the actions of an eavesdropper).  After distillation,
Alice makes a measurement on her half of the singlets, and sends
the results to Bob, who can then transform his half of the singlets
into the state which is to be transmitted.  Since the outcomes of
Alice's measurements are completely random, Eve gets no information
about the state which is being sent.  Teleportation therefore, automatically
ensures encryption.  If the classical channel is authenticated,
teleportation also ensures authentication of the quantum information.
Additionally, the unknown state is never held by Eve,
and so, it cannot get destroyed or corrupted.  

Both teleportation and \qde therefore need one usage of the 
quantum channel for every qubit which is to be sent and encrypted.
The amount of classical communication needed for teleportation $n$ qubits
is one round of $O(\log n)$ classical bits (c-bits) from Bob to Alice,
and $2n$ c-bits from Alice to Bob.  The first round is used in the
distillation process (Bob needs to tell Alice which singlets he tested
for purity, and how noisy the singlets are),
and the second round is needed for Alice to tell Bob the result of her
measurement.  The PQC does not need these two rounds of classical communication,
but it does need the private key.  It therefore seems that this saving
in classical communication for \qde comes at an expensive price (the
private key). 

One might argue that the private key can be recycled, and therefore, it
doesn't represent much of a cost.  However, some of the private
key will be lost 
in the case of heavy eavesdropping or noise.  Furthermore, 
recycling the key also requires interactive communication.  Bob
must inform Alice how much noise is in the check-bits and which
privacy amplification function they will use.  This communication can
occur at a more convenient time (after Bob has decrypted the states),
but it is still needed.

We therefore inquire into situations where one might prefer the PQC over
teleportation.  One situation where the two are equivalent, is if one wishes
to store the quantum information in one's own lab, but does not trust other people
in the lab.  One could then q-encrypt the states, and keep the key
somewhere safe.  One could also use teleportation, and keep the results of
the measurement locked away, but preparing singlets to teleport a state to
oneself seems rather silly.   Here, we show other applications of the PQC.
In Subsection \ref{sec:sharing} we consider a situation which
we call secure secret sharing which requires either a modified version of \qde or of
teleportation.  In this case, both protocols require the same amount
of resources -- namely, a private key and 2 cbits of communication.
We also find the interesting
result that in this case, the private key, which is used to encrypt
classical information, can also be recycled. We will present an asymmetric
version of secure secret sharing (one party is more trusted) in which the
teleportation protocol is more resource intensive. We also explore two
other examples where one wouldn't use teleportation, namely, the
protection of quantum resources such as entanglement or the memory
of a quantum computer.  
For protecting entanglement, the private key is half the size of the 
general case.
These examples are discussed in Subsection
\ref{sec:entanglement}.  It is the the protection
of quantum information where \qde may find its most useful application.

\subsection{Encrypted quantum secret sharing}
\label{sec:sharing}
Let us first consider the following scenario which we will call encrypted quantum
secret sharing.  We imagine a scenario similar to
original secret sharing\cite{secret-sharing}:
Alice wants to send a quantum state to Bob
and Claire, but she does not want either one to be able to use the state
separately. Alice 
desires that they cooperate in order to decrypt the state. 
However, we also imagine that both
Bob and Claire are extremely adversarial and may try
to eavesdrop on each other's communication channels in order to learn the
state.  Or perhaps, there is a concern that another party may eavesdrop on
the channel.

Consider the following modification of the QAS 
(e.g. Protocol \ref{pro:sqas}).  Alice shares
a classical private key $X$ with Bob (to be used for q-encrypting), 
but rather than q-encrypting the
state using
this key, she instead q-encrypts the quantum state using another random
string $J$ and then applies the authentication protocol using key $S$ (shared
with Claire).
She then sends the encrypted quantum state to Claire.
Since Bob doesn't know $J$, he will gain nothing by eavesdropping on
the quantum channel between Alice and Claire.  Once Claire receives
the quantum state, she checks to see that the
state has arrived intact using $S$.
If the authentication
test fails, they abort.  If the test succeeds, Alice
classically encrypts the random string $J$ using
$X$ and sends the encrypted string to Bob.  This prevents Claire from
learning the key used to encrypt the state she has, while Bob can decrypt $J$
using $X$.   Claire now possesses the encrypted quantum state, while Bob possess
the key, and they will not be able to decrypt it unless they cooperate.  The
total protocol has consumed $2n$ bits of shared private key and $2m$ bits
of classical communication between Alice
and Bob,
and $n$ uses of the quantum channel
from Alice to Claire.
The state can only be decrypted if Bob and Claire get together and
cooperate.

The protocol involving teleportation requires the same amount of resources.
Alice sends half-singlets to Claire, and they use a purification
testing protocol over an authenticated classical channel to ensure that they indeed
hold singlets. Alice then makes the joint measurement on her half-singlets and
the state to be shared.  Alice then encrypts the results of the measurements
using the key shared between herself and Bob, and sends it to him.

Key recycling can be done in both these cases,
although it is a bit more problematic if one is concerned not only
about a party learning the key, but also, correlations between different
states.  The transmitted states will
need two layers of authentication.  Essentially Claire needs
to be sure that either the singlets or the transmitted state
is authentic, and then Bob would need to test for authenticity
when they decrypt.  This would need to be done Claire's lab
and so the situation may be a bit awkward for this to occur 
with Claire hovering over
his shoulder (however, one may assume that they are polite enough to each
other in person to allow this to happen).  Alternatively,
Claire could test for authenticity in the presence of Bob, assuming
that Bob is allowed to test Claire's apparatus.  Key recycling makes
more sense in the case where one is concerned about eavesdropping from
some other party, in which case only one layer of authentication is needed,
since Alice can trust Claire's acceptance of the authentication.

This recycling in the case of encrypted quantum secret sharing is perhaps more surprising
than recyclability in the original PQC protocol, since in this case,
the information being encrypted is classical.  In the case of teleportation,
it is the result of Alice's measurements which, although random, could be
used by Claire to learn what state she has.  The proof of this recyclability
is almost identical to Theorem \ref{thm:main}.  The only difference is that
here, one does not need to show that Eve must be product with the message
($J$ in the case of QAS-type protocol, and the result of Bell measurements
in the case of the teleportation protocol).  
Indeed, she will not be, as both strings are classical.  However, we do not
need this to be the case, because there are no correlations from one round
to another -- the messages are completely random strings.

One can envision situations where the PQC is less resource intensive
than teleportation.  Consider the following: Alice wants to send a state
to Bob, but because quantum channels and labs are expensive, Bob has
neither of these.  The only option then, is for Alice to send the state
to Claire, a technician in a large company which owns all the quantum
channels, and which also has  storage equipment for keeping quantum states.
Claire is however, not trusted.

Alice can however, encrypt the quantum state using the key shared between
her and Bob, and send the state to Claire.  Bob can then go over to Claire's
lab any time he chooses, and decrypt the state (we once again imagine that
he is able to test the equipment first).  The teleportation protocol
however, uses more resources.  The half-singlets are sent to Claire,
and the results of the measurements sent to Bob.  The measurement results
that are sent
to Bob have to
be encrypted, to ensure that Claire doesn't eavesdrop on Bob's classical
channel.  Therefore, the teleportation protocol also needs
a private key.  It is therefore more expensive, because it needs two
way communication, while the PQC does not.

\subsection{Protecting entanglement}
\label{sec:entanglement}

We now turn to the issue of using PQC to protect various
resources, and we demonstrate a scenario where
only $n$ bits of private key are necessary and sufficient rather than $2n$.
Let us imagine the following:
there is a teleportation
device\cite{teleportation} set-up between Alice (who is on Earth)
and Claire (a robot) who is on the planet Venus.  The more
practical reader can instead imagine a device which can be used
to perform quantum key distribution between Earth and a satellite,
and we note that related devices for satellite deployment are already
being tested\cite{satallite}\footnote{It is usually simpler to
perform quantum key distribution using sent photons, but one can also
use singlets as described in the protocol of \cite{ekert-qkd}.}.
Now, the teleportation device
between Alice and Claire is essentially made up of many entangled
states shared between Alice and Claire which we take to be the singlet
\beq
\psi_-=|01\ra_{AC}-|10\ra_{AC} \s .
\label{eq:singlet}
\eeq
We assume that these singlets have already been authenticated,
and so Alice possesses many singlet-halves which look locally
like maximally mixed states. She may decide that she
will not use the
device, but instead wants to send her states to Bob for his use.
They may have two concerns: 1) to ensure that
an adversary Eve has no idea what quantum states are being
sent, or 2) they may not care what Eve knows but want
to ensure that Eve will not steal the states for her own use
(since she may want to use the device to also teleport something to
Venus).

Here, we consider the latter case (which we shall refer to
as quantum data protection), and assume that Eve
is also located on Earth.  One way to ensure that
the singlets will never fall into the wrong hands is to have
an authentication protocol set up between Claire, and the authorized party on Earth.
However, this is only useful if one can be sure that Claire's singlets
will also never fall into the wrong hands.  It may be that
Eve has a friend on Venus who might take control of the teleporter.

Alice, would therefore want to perform some sort of protection protocol
on her half of the singlets, to ensure that if Eve steals them,
they will be of no use.  In this case, we do not need the
full encryption scheme of \cite{boykin-qe,mosca-qe} which uses $2n$ bits of private key,
but can instead
use a scheme using only an $n$ bit private key.  We now show
that $n$ bits are necessary and sufficient to protect entanglement.

To show sufficiency, consider the following protocol.
Alice and Bob share a private key $k$
and Alice divides her states into large blocks of size $n$
and performs the bit flip operation ($\sigma_x$) on each of her qubits
conditional on each bit of the private key.
The total state of each qubit is then
\beq
\rho_{ABC}=\frac{1}{2}(|00\ra \la 00|\otimes\ket{\psi_-}\bra{\psi_-}
+|11\ra \la 11|\otimes\ket{\psi_+}\bra{\psi_+} )
\eeq
where $\psi_+$ is the Bell-state $|00\ra_{AC}+|11\ra_{AC}$ shared between
Alice and Claire, and
the states $|00\ra$ and $|11\ra$ is the classically
correlated private key shared between Alice and Bob.
An adversary who does not have access to the key, receives
the state
\beq
\rho_{CE}=\frac{1}{2}|\psi_-\ra\la\psi_-|
+\frac{1}{2}|\psi_+\ra\la\psi_+|
\eeq
It is immediate to find using e.g. the partial transpose test 
that this state is completely unentangled,
and is of no use whatsoever to an adversary wishing to
use the singlets.

To show that $n$ bits of private key are necessary, we note
that we would not be able to use less than $n$ bits of randomness
because any mixture of a singlet with another pure state is
unentangled unless p=1/2.  More explicitly,
the encryption amounts to applying some unitaries $U_i$
with probabilities $p_i$ to Alice's halves of singlets.
The amount of used bits is then given by the entropy
 $H(\{p_i\})$. To get a bound for the entropy, we note
 that Alice wants the state to be separable, while
the separable state satisfies $S(\varrho_{AB})\geq S(\varrho_A)$
\cite{Hor1994}.
In our case $\varrho_{AB}$ is
the state obtained from singlets after applying the
random unitaries. We have $S(\varrho_A)=n$, where $n$ is amount of singlets while
$S(\varrho_{AB})\leq H(\{p_i\}) $.  If the state is to be separable,
 $H(\{p_i\})$ must be then no smaller than $n$.

After protecting the singlet, Alice can send her $n$ qubits to Bob.
Bob can then
test the received qubits to ensure that Eve has not stolen
any, by simply testing for purity (for example, by using the stabilizer
codes used in Protocol \ref{pro:sqas}).
In the above scenario, we are not encrypting quantum information
but rather protecting it -
we assume that Eve already knows that Alice is trying to send
the half-singlets to Bob.

We can use key-recycling also in this case.  If, for example,
we use the additional layer of authentication, and the test passes,
then Alice and Bob reuse the key.  Here, they can reuse the key,
even though it is only $n$ bits long to begin with.  This is
because the bound given in the Appendix ensures that Eve can
learn nothing about the key.
The half singlet states are already maximally
mixed, and so there is no measurement that Eve can make which would allow
her to guess the private key.

Finally, one can also use quantum data protection even when not sending
data to another party.  One might imagine that Alice is concerned that
someone in her lab (Eve) might steal the entanglement and use it for some
unauthorized purpose.  Alice could use quantum authentication, which would
let her know that the entanglement has been tampered with, however by then
it is too late, and Eve has possession of the half-singlets.  Again, setting
up an authentication protocol is no good if Eve has a friend on Venus who
might also steal the other half of the singlets.  Alice can therefore use
the protocol above using her own private key.  This private key can easily
be stored somewhere safe, as opposed to the quantum states which presumably
must be stored in a prominent place in a lab, as the states need to constantly
be protected against decoherence.  Note that in the case of storage, a
teleportation protocol makes little sense.


There are other example's where one can use the PQC to protection
quantum data.  For example, one might one to protect the
memory of a quantum computer, especially if one is running a long
factoring algorithm, and one doesn't trust other people in the lab.  
In such a case, one might make do with a key which is only as long
as the threshold, beyond which error correction is
impossible\cite{dorit-bo-thresh,shor-fault}.  I.e.
one introduces errors conditional on a key, such that the computation
can no longer be performed.  These errors can be undone if one has the
key, and  the computation can proceed, but without the key, the computer
will not run.  Whether an adversary might still be able to get some
information by performing some measurement on the quantum memory in such a scenario is not clear.


\section{Discussion}
\label{sec:conclusion}

>From a conceptual point of view, the private quantum channel and the
authenticated quantum channel are  interesting
because it allows us to decouple the sending of a quantum state, and the
encryption/authentication of a quantum state which are automatically coupled 
in teleportation.  
Since teleportation does not require a private key, we have inquired into
the necessity of the private key used in the QAS.  Indeed, we find that in
some sense, the private key is not needed, as one can keep reusing the same
private key, adding only a small amount of additional private key each time.
The role of the private key was seen to be more significant in terms of
communication -- it allows the QAS to be non-interactive.  It would be
interesting to better understand the role of the private key.  If it can be
recycled, one wonders the extent to which it is needed at all.

Encryption of quantum information is different from encryption in the
classical case.  One can have the exact opposite scenario: Instead of Alice
knowing the state, and Eve trying to learn it, one will likely be 
concerned about
the case of Alice not knowing the state, and Eve knowing something about
it.   We therefore introduced the notion
of {\it protecting} quantum resources.  Namely, the private quantum channel
allows us to ensure that certain resources cannot be used by an adversary.
This is in contrast to the usual role of encryption, which seeks to prevent
an adversary from knowing a message.  It was found that for protecting some
resources (such as entanglement), less private key was needed.  Other
applications, such as an encrypted version of quantum secret sharing were also
introduced.

We also introduced a basic law of privacy which governs how the size of the
private key changes as a function of sent qubits and encrypted messages.
We believe it would be interesting to explore this law further, as well as
the analogy between eavesdropping and thermodynamics.  

\begin{acknowledgments}
%
This work is supported by EU grant EQUIP, No. IST-1999-11053, and
PROSECCO No. IST-2001-39227.
J.O. also acknowledges the support of the
Lady Davis Trust, and
ISF grant 129/00-1.  We thank Daniel Gottesman, Debbie Leung and
Dominic Mayers for valuable
discussion.  We also thank Michele Mosca for drawing our attention to
reference \cite{debbie-qvc}.
We thank the MSRI for their hospitality
during the 2002 Quantum Computation program while this work was being
conducted.

\end{acknowledgments}


\begin{thebibliography}{30}
\bibitem[{xor()}]{xor}
\bibinfo{note}{An XOR flips a target bit, depending on whether a control bit is
  a 0 or 1. In this case, the target bit is the message to be encrypted, and
  the control bit is the secret key}.
\bibitem[{dis()}]{distinction}
\bibinfo{note}{One might think that if one can recycle the key, that the size
  of the key does noxt matter. However, we will see that this is not the case,
  since encryption must be done in large blocks before each round of recycling.
  Furthermore, in some cases, the key-recycling protocol will be aborted, in
  which case, the key is discarded.}
\end{thebibliography}

\end{document}